This article has been accepted for publication in IEEE Antennas and Wireless Propagation Letters. This is the author's version which has not been fully edited and content may change prior to final publication. Citation information: DOI 10.1109/LAWP.2024.3513403IEEE ANTENNA AND WIRELESS PROPAGATION LETTERS    1# Compact High-Efficiency All-Polarization Rectenna for Wireless Power Transmission

Haoming He, *Graduate Student member, IEEE*, Zongyang Dan, Zhongqi He,
and Changjun Liu, *Senior member, IEEE**Abstract*—In this paper, we present an innovative design for a compact, high-efficiency all-polarization receiving rectenna tailored for wireless power transmission. This rectenna, which integrates an antenna with two same rectifier units, employs direct conjugate matching of antenna impedance to rectifier impedance. This approach eliminates the necessity for an external impedance matching network, thereby reducing the overall dimensions of the rectenna. The implementation of virtual ground concept streamlines the design of the rectifier's output filter. The low-profile antenna, engineered for operation at 2.45 GHz, demonstrates high conversion efficiency across all polarization angles. The measured RF-to-DC efficiency exceeds 63% for all polarization angles, achieving a peak efficiency of 82.2%.

*Keywords*—All Polarization, Conjugate matching, Linear polarization, Rectennas, Wireless power transmission.## I. INTRODUCTION

Microwave wireless power transmission (WPT) technology has garnered significant attention in recent years as a promising solution for next-generation power transmission systems [1][2][3][4]. This technology is particularly advantageous for wireless sensor network systems, as it has the potential to eliminate the need for battery replacement[5][6][7] and to reduce maintenance requirements for Internet of Things (IoT) devices [8][9][10][11]. Within a WPT system, the rectenna—comprising both the antenna and rectifier—plays a crucial role in harvesting incident RF energy and converting it into DC power [12][13]. Given that the polarization of incident electromagnetic waves can be unknown or time-varying in most WPT scenarios, antennas capable of polarization diversity are preferred to maximize RF energy collection for WPT applications.

To enhance the reception efficiency of rectennas, antennas are typically designed to capture linearly polarized (LP) waves, which exhibit high energy reception efficiency [14][15]. However, when a circularly polarized incident wave is encountered, the power received by a linearly polarized antenna can be reduced by as much as 50%. To address this limitation, rectennas have been developed to specifically receive left-handed circularly polarized (LHCP) and right-handed circularly polarized (RHCP) waves [16][17][18]. Despite these advancements, performance remains constrained, prompting the design of antennas capable of dual circular polarization

This work was supported by NFSC under Grant U22A2015 and 62071316. (Corresponding author: *Changjun Liu*.)
The authors are with the School of Electronics and Information Engineering, Sichuan University, Chengdu 610064，and also with the Yibin Industrial Technology Research Institute, Sichuan University, Yibin 644001, China (e-mail: cjliu@ieee.org).[19][20]. Concurrently, the integration of dual-polarization antennas in rectennas to capture fully polarized waves has emerged as a significant area of research [21]-[27]. These designs enable the reception of electromagnetic waves of arbitrary polarization, making them suitable for energy transmission systems and environmental energy harvesting. However, they necessitate more complex matching networks and rectifier units, complicating the design of all-polarization rectennas and increasing the overall dimensions of the antennas

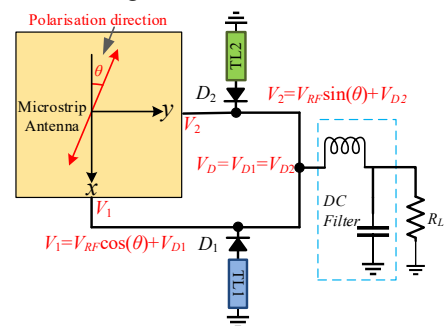

Fig. 1. Schematic diagram of the designed all-polarization rectenna.

This paper presents a compact and straightforward all-polarization rectenna design, utilizing an orthogonal feeding technique for the microstrip patch to achieve dual polarization. The conjugate matching of the antenna impedance with the diode impedance minimizes the complexity of the matching network. Additionally, the implementation of a harmonic suppression structure enhances overall efficiency, allowing received electromagnetic waves of varying polarizations to be processed through a unified direct flow path. The rectenna designed using this approach demonstrates high rectification efficiency across a range of polarization angles.

## II. PRINCIPLE AND DESIGN METHOD

Fig. 1 is a schematic diagram of the designed all-polarization rectenna. Compared with the standard all-polarization rectenna with dual-polarization antennas, the rectifier is directly connected to the antenna, and there is no matching network and low-pass or band-pass filter. For the incident waves with different polarizations angles, the voltage of the input electromagnetic wave is set to $V_{RF}$, and $V_D$ is the DC output voltage. The dc energy converted by electromagnetic waves share a common output port with DC voltage output in parallel without passing through the output filter, $V_D=V_{D1}=V_{D2}$. The voltages at the two ports are $V_1= V_{RF}cos\theta+V_{D1}$, $V_2=V_{RF}sin\theta+V_{D2}$. The impedance matching network and the design of multiple filters are eliminated, reducing the electromagnetic energy loss and making the rectenna's structure simple and compact.

© 2024 IEEE. Personal use is permitted, but republication/redistribution requires IEEE permission.  See https://www.ieee.org/publications/rights/index.html for more information.

## A. Design of Rectenna

A dual-polarization antenna will have two orthogonal modes. In the case of a microstrip antenna fed via microstrip line, two feed ports can be positioned perpendicularly to achieve orthogonal polarization. The feed structure of the designed microstrip patch antenna is shown in Fig. 2. When the amplitude and phase of the feed are matched, two orthogonal modes $TM_{10}$ and $TM_{01}$ are excited simultaneously. This configuration results in the generation of two orthogonally polarized linear waves in space, with strong isolation between the two ports. Ports 1 and 2 serve as feed points that connect to the rectifier diode.

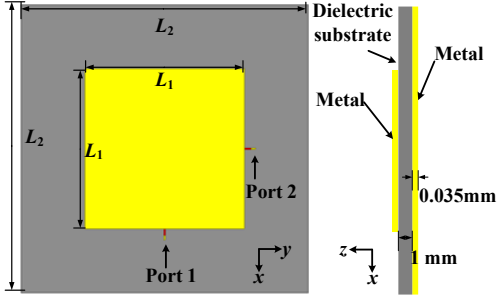

Fig. 2. The structure designed microstrip patch antenna.

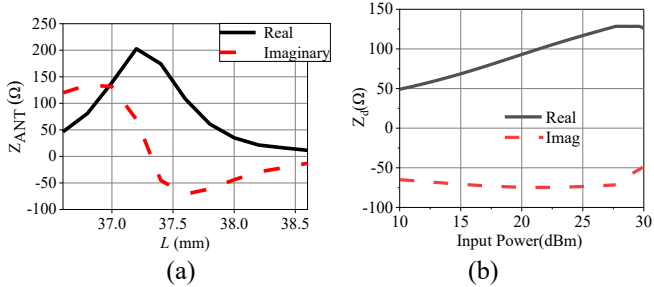

(a)      (b)

Fig. 3 (a) The input impedance of the antenna varies. (b) The input impedance of the diode varies

TABLE I
SPICE PARAMETERS OF AN HSMS282 DIODE

| $B_v$ | $C_{j0}$ | $I_s$ | $R_s$ | $V_{bi}$ |
|---|---|---|---|---|
| 15 V | 0.7 pF | 0.022 μA | 6.0 Ω | 0.25 V |

The initial step involves determining the dimensions of the antenna patch. Choosing a square patch as the geometry of the dual-polarized microstrip antenna, we can estimate the size of the square patch using a formula, and the calculation process is as follows

$$\lambda_g = \frac{\lambda_0}{\sqrt{\frac{\varepsilon_r+1}{2}+\frac{\varepsilon_r-1}{2}\left(1+10\frac{h}{w}\right)^{-\frac{1}{2}}}} \quad (1)$$

$$L = \lambda_g / 2 \quad (2)$$

$\lambda_0$ and $\lambda_g$ are the operating wavelength and guide wavelength at the operating frequency $f_0$, $\varepsilon_r$ and $h$ are the relative dielectric constant and thickness of the substrate, respectively, and $w$ is the width, respectively. For dual-polarization microstrip antennas, the lengths of the sides are equal.

We fabricated the patch with an F4B-2 double-sided copper-clad board with a relative permittivity of 2.65, a loss angle tangent of 0.005, and a thickness of 1 mm. When the operating frequency $f_0$=2.45 GHz is set, the dimensions is optimized. The edge length of the antenna is varied, and the corresponding input impedance is measured, as illustrated in Fig. 3(a). The antenna size can be adjusted to match the input impedance of the diode.

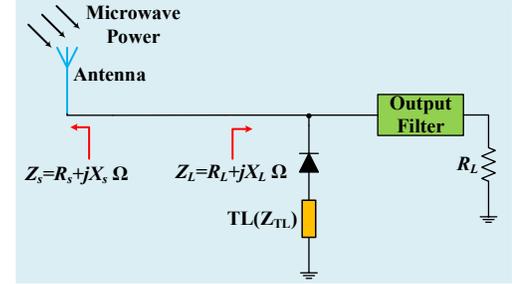

Fig. 4. Structure diagram of a compact rectenna for a new impedance-matching network

## B. Rectifier Design

The newly designed dual-polarization rectenna employs a full-wave simulation to obtain the impedance characteristics of the microstrip patch antenna, identifying the position that achieves conjugate matching with the Schottky diode. Subsequently, the diode is directly connected between the patch and the ground of the antenna. This approach eliminates the need for additional microstrip branch lines for impedance adjustment. The design schematic of the rectifier is presented in Fig. 4.

The Schottky diode is an essential part of a rectenna and converts microwave to dc power. The impedance of a Schottky diode is

$$Z_D = \frac{\pi R_s}{\cos\theta_{on}\left(\frac{\theta_{on}}{\cos\theta_{on}}-\sin\theta_{on}\right)+j\omega R_s C_j\left(\frac{\pi-\theta_{on}}{\cos\theta_{on}}+\sin\theta_{on}\right)} \quad (3)$$

where $R_s$ is the series resistance of the diode, $\theta_{on}$ is the turn-on angle of the diode, and $C_j$ is the junction capacitance of the diode. The SPICE parameters of an HSM282 are shown in Table I. The input impedance of the diode with 220 Ω load under ideal conditions is analyzed, as demonstrated in Fig. 3(b).

In order to improve the rectifying efficiency of the rectifier, various harmonic suppression structures have been integrated into the design. The adopted harmonic suppression structure in series with the diode does not impact the antenna's performance or dimensions and can improve the rectifying efficiency of the rectenna.

In Fig. 4, TL is $\lambda_g/8$ short-ended transmission lines at $f_0$. A rectifier with a $\lambda_g/8$ short-ended transmission line has been investigated in our previous work [28], and its input impedance is given by

$$Z_{TLi} = jZ_{TL}\tan\left(\frac{\pi}{4}\frac{f}{f_0}\right) = \begin{cases} 0 & f=0 \\ jZ_{TL} & f=f_0 \\ \infty & f=2f_0 \\ -jZ_{TL} & f=3f_0 \end{cases} \quad (4)$$

where $Z_{TL}$ is the characteristic impedance of TL. The input impedance exhibits inductive characteristics, compensating for the capacitive reactance of the diode. This arrangement facilitates impedance matching and aids in recycling the second harmonic, thereby enhancing rectification efficiency

Consequently, TL can adjust the imaginary impedance of the diode. Variations in antenna size will affect the real part of the antenna impedance; Then conjugate matching of the rectifier and antenna can be realized by joint debugging of the two.

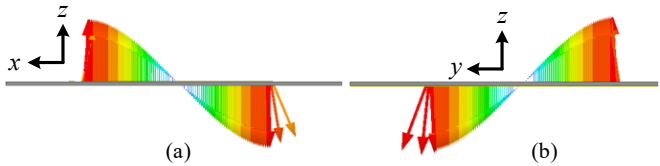

Fig. 5. The electric field distribution of the excitation antenna distributed on both ports. (a) port 1;(b) port 2.

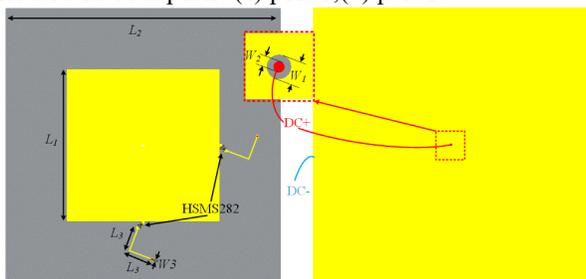

Fig. 6. All-polarization microstrip rectenna layout.

TABLE II
DIMENSIONS OF PATCH ANTENNA

| Dimension | $L_1$ | $L_2$ | $L_3$ | $W_1$ | $W_2$ | $W_3$ |
|---|---|---|---|---|---|---|
| Value(mm) | 37.6 | 67.6 | 5.8 | 1.8 | 0.7 | 0.2 |

At an input power of 24 dBm, the input impedance of the diode is 81-$j$86 Ω. Therefore, by adjusting the size of the antenna, the side length $L$ of the patch antenna can be 37.7mm, and the input impedance of the antenna is 81-$j$65 Ω, which satisfies $Re(Z_{ANT})=Re(Z_d)$. Subsequently, the input impedance of the harmonic suppression structure TL, i.e., the width of the $\lambda_g/8$ short-ended transmission line, is adjusted to make its characteristic impedance $Z_{TL}= -Im(Z_d+Z_{ANT})$. $Z_{TL}$=151 Ω was calculated. According to the microstrip theory, the width of the corresponding microstrip line is 0.19 mm. Folding the microstrip line allows it to be achieved within the limited dimensions of the antenna. In order to realize the all-polarization rectenna, the same rectifier units are connected to ports 1 and 2 of the patch antenna, respectively

*C. dc Filter Design*

Regarding the DC output of the rectenna, this paper utilizes the concept of virtual ground to establish the DC output path. This is achieved by analyzing the electric field distribution resulting from the excitation of the two ports independently, as well as identifying the virtual ground position at the intersection point of minimal electric field intensity. As illustrated in Fig. 6, the electric field distribution when both ports are energized reveals that the center of the patch serves as a common virtual ground for the two modes. Consequently, in designing the DC output path, a small section of the substrate is exposed on the ground plane through a hole at the center of the square patch. This configuration establishes the DC output path without adversely affecting the radiation characteristics of the microstrip antenna. Currently, the microwave energy of the all-polarization antenna does not enter the dc load, and the microwave signal is effectively blocked; no additional design is required. At the same time, the output energy of the two rectification units corresponding to all the polarized antennas is output to the same output load in parallel [29] , which helps the rectenna to receive electromagnetic wave energy in multiple polarization directions and improve the effect.

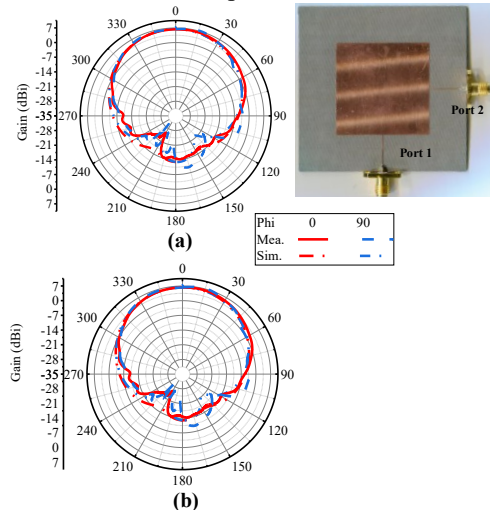

Fig. 7. Antenna direction map（a）port 1; (b) port 2.

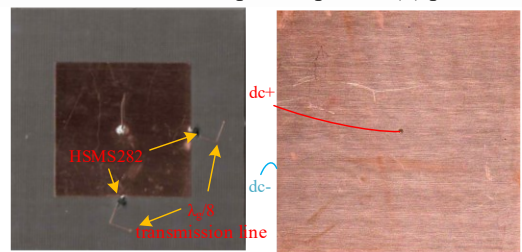

Fig. 8. Photograph of the all polarized rectenna processed.

Following the co-simulation and optimization adjustments of the rectenna, the configuration of the rectenna is presented in Fig. 6. The parameters of the all-polarization microstrip rectenna are summarized in Table II.

III. IMPLEMENTATION AND MEASUREMENT

Firstly, the input impedance of the two ports is matched to 50 Ω under the exact size of the patch, and the gain of the dual-polarized microstrip antenna is tested, which provides data for the subsequent measurement of the rectification efficiency of the rectenna. Fig. 7 shows a photograph of a microstrip patch antenna without a rectifier with the simulated and measured gain at φ=0° and φ= 90° of the proposed rectenna, which present a good agreement. According to the test results, the maximum gain of the fabricated dual-polarized microstrip antenna is 6.4 dBi.

Fig. 8 is the front and back photos of the processed orthogonal polarization microstrip rectenna. Two HSMS282 diodes are distributed and connected to Port 1 and Port 2, and the grounding through hole is designed in the middle of the patch as a dc loop, which dc output from dc+. Fig. 9 is a schematic diagram of the measurement of an all-polarization rectenna. The microwave signal output by the microwave source (Agilent E8267C) is amplified by the power amplifier and then divided into two parts in the directional coupler; one part is used for the power meter (Eanalyzer AV2433) to detect the output power of the microwave, and the other is transmitted through the standard horn antenna. The rectenna is tested in the far field ($R$=1 m) simulation of the transmitting antenna, followed by a standard resistance box and voltmeter (FLUKE multimeter) to absorb dc power and detect the output voltage.

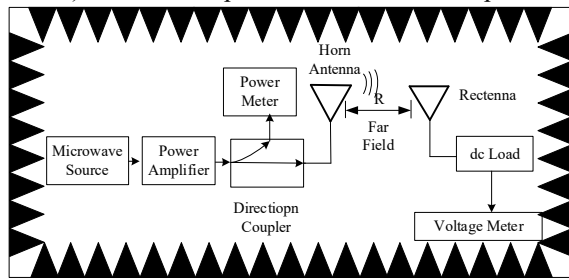

Fig. 9. The schematic diagram of the measurement.

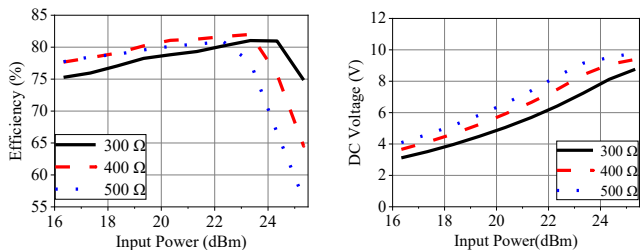

Fig. 10. (a) efficiency and (b) output voltage with variation of input power under different loads.

The formula for calculating the conversion efficiency of the proposed rectenna is as follows:

$$\eta_{RF-dc} = \frac{P_{dc}}{P_{IN}} \times 100\% = \frac{V_{OUT}^2}{R_{dc}} \times \frac{1}{P_{IN}} \times 100\% \quad (5)$$

where $P_{IN}$ is the input power of the rectifier, $P_{dc}$ is the output dc power, $R_{dc}$ is the dc load, and $V_{OUT}$ is the output dc voltage.

For all-polarization rectennas, performance needs to be tested in all receive directions. The polarization angle $\theta$ represents the polarization angle of the transmitting antenna, that is, the linear polarization direction of the electromagnetic wave received by the orthogonally polarized antenna. At a load of 400 Ω and a power of 23.3 dBm from the input rectifier, the highest rectification efficiency of 82.2% was obtained, as shown in Fig. 10 (a). Fig. 10 (b) shows the output voltage as a function of power under different loads.

Subsequently, to test the reception performance in all polarization directions, different polarization angles were rotated. A transmit power of 23.3 dBm was selected for testing. The rectification efficiency and output voltage at various angles are shown in Fig. 11. According to the test results, it can be seen that at a fixed load of 400 Ω, a voltage higher than 5.2 V can be obtained in all different directions of polarization waves and a rectification efficiency of more than 63% can be achieved. For a given received power, the load value can be reduced to obtain greater rectification efficiency. Both rectifiers are in a similar state at ±45° polarization incidence. According to the parallel output theory[29], the best efficiency is obtained when the load drops by half to 200 Ω.

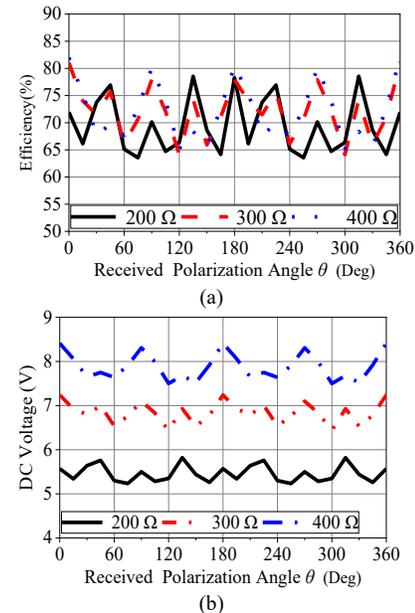

Fig. 11. The rectification efficiency (a) and output voltage (b) at different angles.

Comparisons with other all-polarization rectenna are shown in Table III. The design method proposed in this paper has more obvious advantages than other rectennas regarding integration and rectification efficiency.

TABLE III
COMPARISON WITH PREVIOUS RECTENNAS

| Ref | [23] | [24] | [25] | [27] | This work |
|---|---|---|---|---|---|
| Frequency (GHz) | 2.45 | 2.45 | 2.45 | 1.9 | 2.45 |
| Maximum efficiency | 78% | 52.3% | 69% | 62% | 82.2% |
| All-polarization min efficiency | 61% | 44.3% | N.A | 35% | 63% |
| Antenna Structure | Multilayer patch antenna | Patch antenna | Multilayer patch antenna | 3D | Patch antenna |
| Separate rectifiers | Yes | Yes | Yes | Yes | No |
| dc passage | 1 | 4 | 2 | 1 | 1 |

## IV. CONCLUSION

A compact, high-efficiency all-polarization rectenna was designed, analyzed, and processed. The conjugate design of antenna impedance and rectifier diode impedance removes the matching network. The virtual ground concept eliminates the output filter design, contributing to a more compact rectenna design. The band-stop structure can be used to adjust the impedance of the imaginary part of the diode and enhancing the efficiency of rectification. The rectenna has been measured to have an efficiency of more than 63% over the entire polarization plane, with a maximum efficiency of 82.2%. This simple and efficient design method can be used in designing wireless power transmission systems.